# *k*-DAG Based Lifetime Aware Data Collection in Wireless Sensor Networks

Jingjing Fei[1], Hui Wu[1] and Yongxin Wang[2]

[1]School of Computer Science and Engineering, The University of New South Wales, Sydney, Australia

[2]FEIT, University of Technology, Sydney, Australia

## ABSTRACT

*Wireless Sensor Networks need to be organized for efficient data collection and lifetime maximization. In this paper, we propose a novel routing structure, namely k-DAG, to balance the load of the base station's neighbours while providing the worst-case latency guarantee for data collection, and a distributed algorithm for construction a k-DAG based on a SPD (Shortest Path DAG). In a k-DAG, the lengths of the longest path and the shortest path of each sensor node to the base station differ by at most k. By adding sibling edges to a SPD, our distributed algorithm allows critical nodes to have more routing choices. The simulation results show that our approach significantly outperforms the SPD-based data collection approach in both network lifetime and load balance.*

## 1. INTRODUCTION

A WSN (Wireless Sensor Network) consists of a set of sensor nodes. A sensor node is composed of sensors, a processor, wireless communication components and a power module. All the senor nodes in a WSN are connected wirelessly, and work cooperatively to send the sensed data to a base station. The size of a WSN varies with applications. In a smart home, a WSN may have just dozens of sensor nodes. In a bushfire detection application, the area covered by a WSN may span several square kilometres with thousands of sensor nodes deployed. In some applications such as border surveillance, data need to be collected in real-time. Therefore, it is desirable to minimize the maximum latency of data collection, i.e., the maximum time taken by any message to arrive at the base station from the source sensor node.

In WSNs, sensor nodes are typically battery-powered, and usually deployed over a large area or in a hostile environment, which makes frequent battery replacement impractical. As a result, optimizing the energy consumption of sensor nodes is critical for extending the network lifetime. Typically, wireless communication consumes most energy of a sensor node, compared to computation and sensing [1-3]. Therefore, lowering the energy consumption of wireless communication can significantly save sensor nodes' energy, increasing the lifetime of a WSN. The communication range of a sensor node is constrained by the transmit power. To save energy, the transmit power is kept low, leading to a short transmission range. As a result, the communication between data source nodes and the base station is commonly achieved in multi-hop way. Therefore, the routing topology has a significant impact on the network lifetime.

To prolong the network lifetime, various topologies and routing algorithms have been proposed. Trees are easy to construct without much protocol overhead, and they are widely used in WSNs. In a tree, all the data converge to the base station. For each sensor node, there is only one path reaching the base station so that routing algorithms are easy to implement. However, trees are not robust enough. A link failure caused by any sensor node may isolate all its descendants from the network. Furthermore, the nodes closer to the root are more likely to die sooner as they need to relay more messages from their descendants to the root.

DAG has been proposed to improve the robustness of communication. It is more robust than a tree as each node in the network may have more than one path to the root. In addition, a DAG achieves better load balance than a tree as there are multiple paths from each source node to the base station, resulting in a longer network lifetime. Mesh network is the most robust topology. However, it induces more intricate routing algorithms than a simple tree.

In this paper, we study the problem of lifetime and latency aware data collection in a static WSN where the locations of all the sensor nodes are fixed and there is only one base station. Our objective is to maximize the network lifetime while providing the worst-case latency guarantee. The lifetime of a WSN is defined as the time when the first sensor node dies. We propose a distributed algorithm to construct a $k$-DAG. Our distributed algorithm constructs a $k$-DAG from a SPD (Shortest Path DAG) [4] by adding sibling edges. We make the following major contributions:

- We propose a novel routing structure, namely $k$-DAG, which can improve the lifetime and the robustness of a WSN while providing the maximum latency guarantee.

- We propose a distributed algorithm for constructing a distributed $k$-DAG.

- We propose a novel scheme for naming sensor nodes to support efficient point-to-point routing.

- We have simulated our approach and the approach proposed in [4]. The simulation results show that our approach outperforms theirs by up to 82% in terms of network lifetime.

- As far as we know, our approach is the first one that aims at maximizing the lifetime of a WSN while providing the maximum latency guarantee.

The rest of the paper is organized as follows. Section 2 overviews the existing approaches to lifetime aware routing. Section 3 describes our distributed algorithm for constructing a $k$-DAG. Section 4 presents our simulation results and analyses. Section 5 concludes the paper.

## 2. RELATED WORK

Lifetime aware data collection is a critical issue in WSNs. Different energy consumption models of sensor nodes have been presented and analysed, and a large number of approaches to lifetime aware data collection have been proposed.

[1] proposes a fundamental energy consumption model for sensor nodes. It considers the impacts of both the hardware and external radio environment of sensor nodes. [3] presents a realistic energy consumption model which identifies the energy consumption of each part of the sensor node and the impact of the external radio environment. The power consumption for receiving data is modelled as a constant value. For transmitting, only the power consumed by the power amplifier varies with the transmission range $d$ while the power consumed by the other parts is a

constant. Based on the analyses and the simulation results, it shows that the single hop routing is always more energy efficient than multi-hop routing when a target is single hop reachable. This conclusion encourages the use of greedy approaches to resolve energy efficient routing issues in WSNs.

SPT (Shortest Path Tree) is a commonly used topology in WSNs as each sensor node in a SPT reaches the root with the smallest number of hops. However, a randomly constructed SPT may not increase network lifetime. [5] proposes a new weighted path cost function improving the SPT approach. In this approach, each link is assigned a weight according to its path length to the root, and a link closer to the root has a larger weight. By balancing load according to the links' weights, this approach increases network lifetime compared with those randomly constructed SPT. [6] studies the problem of finding a maximum lifetime tree from all the shortest path trees in a WSN. They first build a fat tree which contains all the shortest path trees. Then, they propose a method based on each node's number of children and its initial energy to find a minimum load shortest path tree to convert the problem into a semi-matching problem, and solve it by the min-cost max-flow approach in polynomial time. [7] proposes an approximation algorithm for maximizing network lifetime by constructing a min-max-weight spanning tree, which guarantees the bottleneck nodes having the least number of descendants. The approximation algorithm iteratively transfers some of the descendants of the nodes with the largest weight to the nodes with smaller weights.

[8] studies the load balancing problem in grid topology. It focuses on the energy consumption of the nodes which can communicate with the base station directly. As mentioned above, increasing the lifetimes of these nodes will prolong the network lifetime in most circumstances. The algorithm first builds a tree by absorbing the nodes which have the greatest load to the lightest branches to achieve the initial load balance. Then, it rebalances the tree by moving nodes from the branches with the heaviest load to the neighbouring branches with lighter load. The simulation results show that the routing trees constructed by their algorithm are more balanced than the SPT constructed by Dijkstra's algorithm.

Trees are not robust enough since each node has only one path to the base station. The topology needs to be periodically reconstructed to avoid network disconnection. SPD has been proposed to solve the robustness problem. In a SPD, each sensor node may have more than one parent. Multiple paths from each sensor to the base station increase not only robustness, but also network lifetime. [4] considers the issues of balancing the load to achieve longer network lifetime by routing on a SPD. It proposes a modified asynchronous distributed breadth-first search method that is similar to Frederickson's algorithm [9], but without the centralized synchronization between level expansions, to build a SPD. It also proposes MPE (Max-min Path Energy) and WPE (Weighted Path Energy) routing algorithms based on SPD.

[10] proposes a routing mechanism which takes advantage of siblings based on the DAG specified by Routing Protocol for Low-power and Lossy Networks (RPL) from IETF ROLL Working Group [11]. The authors present a detailed rank computation function to avoid loops in a DAG, which satisfies the policy of RPL draft. Then, they propose a routing method which allows no more than one sibling-hop per rank in the DAG to preserve the connection of the whole network while preventing loops in routing.

## 3. $K$-DAG CONSTRUCTION

We aim at maximizing a WSN's lifetime by balancing the load among the base station's children as these nodes are the critical ones for network lifetime. Meanwhile, we provide the worst-case latency guarantee for message delivery. Specifically, we ensure that each message from a source sensor node $v_i$ does not travel more than $k+D_{v_i}$ hops to reach the base station, where $D_{v_i}$ is the minimum number of hops from $v_i$ to the base station, and $k$ is a fixed natural number. In this paper, the lifetime of a WSN is defined as the time when the first node depletes its energy.

## 3.1. Network Model

We assume that there is only one base station in the WSN. All the sensor nodes in the network are static. The wireless communication is reliable, and there is no packet loss or retransmission. All the sensor nodes in the network have the same transmission range and the same initial energy level. The base station has unlimited energy. Each sensor node generates one unit data per time unit.

We define a WSN as an undirected graph $G=(V, E)$, where $V$ and $E$ represent the set of sensor nodes and the set of edges denoting communication links, respectively. There are $n$ sensor nodes in the WSN. Each sensor node is denoted by $v_i$ ($i=1, 2, …, n$). Especially, $v_0$ denotes the base station. An edge $e_{ij}=(v_i, v_j)$ exists in $E$ only if $v_i$ and $v_j$ can communicate with each other directly. The graph $G$ is called connectivity graph. We assume that $G$ is connected. Each sensor node has no knowledge of other sensor nodes in the network at the network initiation stage.

A spanning DAG of $G$ is a DAG for data collection satisfying the following constraints:

- The base station is the only source node.

- Each sensor node sends its data only to its parents.

- For each sensor node $v_i$, there is a directed path from the base station to $v_i$.

A SPD is a spanning DAG of $G$ such that for each sensor node $v_i$, each path from the base station to $v_i$ is a shortest path.

A $k$-DAG is a spanning DAG of $G$ such that for each sensor node $v_i$, the lengths of any two paths from the base station to $v_i$ differ by at most $k$.

Given a spanning DAG and a sensor node $v_i$, the DAG rooted at $v_i$ is a subgraph of the spanning DAG where the set of nodes includes $v_i$ and all the nodes reachable from $v_i$, and the set of edges contains all the edges reachable from $v_i$.

## 3.2. SPD and SPT Constructions

Our approach needs to construct a SPD and a SPT at the beginning. The SPD is used to construct a $k$-DAG, and the SPT is used for efficient point-point communication.

A SPD can be constructed by using the algorithm proposed in [4] which employs the relaxation technique proposed in [12]. A SPT can be constructed from a SPD by selecting only one parent for each sensor node.

## 3.3. Naming

We propose a distributed naming algorithm to assign a unique ID to each sensor node. With these IDs, the base station is able to send a message to any node without flooding. The naming is based on a SPT of the network. The ID of each sensor node is a natural number between $1$ and $n$, where $n$ is the total number of sensor nodes in the network. The ID of the base station is $0$.

Given a subtree $T$ and a set of consecutive natural numbers between $m$ and $m+size(T)-1$, where $m$ is a natural number and $size(T)$ is the number of nodes in the tree $T$, the ID of each node in $T$ is recursively defined as follows.

- The ID of the root of T is $m$.

- F or each child $v_i$ ($i=1, 2, ..., k$) of the root of $T$, the ID of each sensor node in the subtree rooted at $v_i$ is a natural number between $m_i$ and $m_i+size(T_i)-1$, where $size(T_i)$ is the number of nodes in the subtree $T_i$ rooted at $v_i$, and $m_i$ is defined as follows.

    1. $m_1$ is equal to $m$.

    2. For each $i$ ($i > 1$), $m_i$ is equal to $m + \sum_{j=1}^{i-1} size(T_j)$ .

Intuitively, the ID of each sensor node is its rank in the depth-first search order of the SPT. However, distributed depth-first search is slow. The above definition underpins a faster distributed algorithm for implementing our naming scheme.

Our distributed naming algorithm consists of three phases. In the first phase, the base station initiates a message informing each sensor node $v_i$ to compute the size of the subtree rooted at $v_i$. This message is sent to each sensor node in the network. In the second phase, starting from the leaf nodes, each sensor node $v_i$ calculates the size of the subtree rooted at $v_i$ after receiving the sizes of the subtrees rooted at its children. In the third phase, starting from the children of the base station, each sensor node assigns a unique ID to itself. Our algorithm uses the following messages.

- CALCULATE-SUBTREE-SIZE($v_i$). This message is used to inform each sensor node $v_i$ to calculate the size of the subtree rooted at $v_i$.

- SUBTREE-SIZE($v_i$, $size_i$). After each sensor node $v_i$ calculates the size $size_i$ of the subtree rooted at $v_i$, it sends this message to its parent in the SPT.

- ASSIGN-ID($v_i$, $min\text{-}id$, $max\text{-}id$). This message is initiated by the base station and sent to each sensor node $v_i$. $min\text{-}id$ and $max\text{-}id$ are the smallest ID and the largest ID, respectively, of all the sensor nodes in the subtree rooted at $v_i$.

The details of our algorithm are shown in pseudo code in Algorithm1.

## Algorithm 1: Naming

**For the base station** $v_0$:

**for** each child $v_i$

    send CALCULATE-SUBTREE-SIZE($v_i$) to $v_i$

**end for**

$size_0 = 0$

**for** each child $v_i$

    receive SUBTREE-SIZE($v_i$, $size_i$) from $v_i$

**end for**

*min-id =1*

**for** each child $v_i$

    *max-id= min-id + size_i -1*

    send ASSIGN-ID($v_i$, *min-id*, *max-id*) to $v_i$

    *min-id =max-id +1*

**end for**

**For each sensor node** $v_i$:

receive CALCULATE-SUBTREE-SIZE($v_i$) from the parent

*size_i =1*

**if** $v_i$ is a leaf node **then**

    send SUBTREE-SIZE($v_i$, *size_i*) to the parent

    receive ASSIGN-ID($v_i$, *min-id*, *max-id*) from the parent

    *ID_i =min-id*    */* The ID of $v_i$ is min-id*/*

**else**

    **for** each child $v_j$

        send CALCULATE-SUBTREE-SIZE($v_j$) to $v_j$

    **end for**

    **for** each child $v_j$

        receive SUBTREE-SIZE($v_j$, *size_j*) from $v_j$

        *size_i =size_i +size_j*

    **end for**

    send SUBTREE-SIZE($v_i$, *size_i*) to the parent

    receive ASSIGN-ID($v_i$, *min-id*, *max-id*) from the parent

    *ID_i =min-id*

    *min-id = min-id+1*

    **for** each child $v_j$

        *max-id= min-id + size_j -1*

        send ASSIGN-ID($v_j$, *min-id*, *max-id*) to $v_j$

        *min-id =max-id + 1*

    **end for**

**end if**

Figure 1 shows an example of our naming scheme, where the natural number beside each sensor node is its ID.

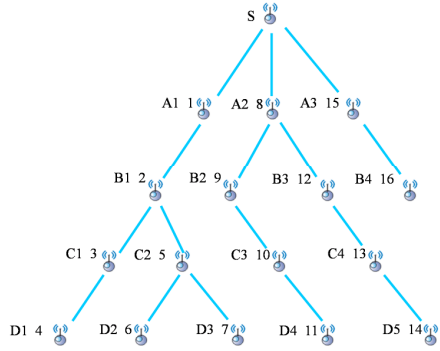

Figure 1. An example of our naming scheme

## 3.3. Load Calculation

After constructing a SPD and a SPT, our approach calculates the load for each sensor node. We use the definition of load in [13] to measure the data flow in the network. The load calculation is based on the DAG constructed so far. Each sensor node calculates its load as a sum of the load it produces and the load coming from its children, and distributes its load to all its parents evenly. Load calculation starts from the leaf sensor nodes in a bottom-up way and ends at the base station. After the base station collects the load from all its children, the load calculation finishes.

**Algorithm 2**: Load calculation

**For each node** $v_i$:

   **if** $v_i$ is a leaf node **then**

      $Ld_{v_i} = 1$

      broadcast LC($v_i$, $Ld_{v_i}$) to all the parents

   **else if** $v_i$ is not the base station **then**

      $Ld_{v_i} = 1$

      **for** each child node $v_j$ **do**

         receive LC($v_j$, $load_{v_j}$) from $v_j$

         $Ld_{v_i} = Ld_{v_i} + load_{v_j}$

      **end for**

      let $p_i$ be the number of parents of $v_i$

      broadcast LC($v_i$, $Ld_{v_i}/p_i$) to all the parents

   **else**  **/\*** $v_i$ is the base station **\*/**

      $Ld_{v_i} = 0$

      **for** each child $v_j$ **do**

         receive LC($v_j$, $load_{v_j}$) from $v_j$

         record the load of $v_j$

**end for**

**end if**

We use an LC message to collect the load information:

- LC($v_i$, $Ld_{v_i}$), where $v_i$ is sender's ID, and $Ld_{v_i}$ is the load flowing from the sender to the receiver.

Algorithm 2 describes in detail how the load of each sensor node is calculated in a distributed way. Figure 2 shows an example illustrating the process of load calculation. Leaf nodes $D_1$, $D_2$, $D_3$, $D_4$, $D_5$, and $B_4$ produce one unit data per time unit to their parents by LC message. When $C_2$ receive all LC messages from its children $D_2$ and $D_3$, it calculates its load which is 5/2 and sends it evenly to its parents $B_1$ and $B_2$. In this way, all the nodes send their load information to their parents, and the load converges to the base station at last. In Figure 2, the value on each edge is the load flowing through the edge.

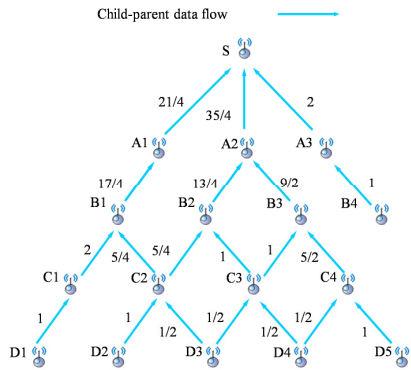

Figure 2. Load calculation based on a DAG

## 3.4. Adding Sibling Edges

The initial $k$-DAG is a SPD. After calculating the load of each child of the base station, our algorithm keeps searching for sibling edges and adding them into the $k$-DAG until the load balance among all the base station children is achieved. At a time, the base station finds a child $v_i$ with the heaviest load, and a child $v_j$ with the lightest load such that there is a sibling edge ($v_t$, $v_s$) satisfying the following constraints:

1. $v_t$ is reachable from $v_j$, but not reachable from $v_i$.

2. $v_s$ is reachable from $v_i$, but not reachable from $v_j$.

A sibling edge ($v_t$, $v_s$) can be added to the current $k$-DAG only iff the following constraints are satisfied:

3. After ($v_t$, $v_s$) is added the current $k$-DAG, the lengths of the longest path and the shortest path from $v_s$ differ by at most $k$.

4. After ($v_t$, $v_s$) is added the current $k$-DAG, the new load of $v_j$ is less than the old load of $v_i$.

A sibling edge $(v_i, v_s)$ will be added if the above constraints are satisfied. Then, our algorithm tries to add adjacent sibling edges which are reachable from $v_i$, but not reachable from $v_j$ before adding sibling edge $(v_i, v_s)$. Once such an adjacent sibling edge is not available, our algorithm will go down to the next level to try to add other sibling edges. This process will be repeated until no sibling edge can be added to the current $k$-DAG or the current node is a leaf node.

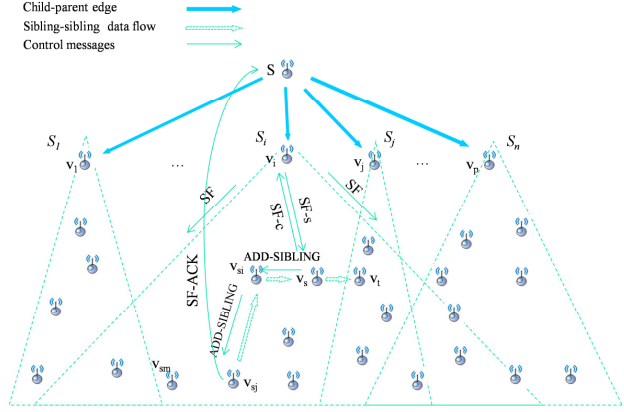

Figure 3. Adding sibling edges

We introduce the following messages:

- SF($v_i$, $v_j$, $LdBl_{v_i}$): The base station floods this message to the DAG rooted at $v_i$ to find a sibling edge $(v_i, v_s)$ as discussed above. $LdBl_{v_i}$ is the ideal load that needs to be diverted from $v_i$ to $v_j$. $LdBl_{v_i}$ is set to $(Ld_{v_i} - Ld_{v_j})/2$.

- SF-c($v_s$, $v_i$, $v_j$, $LdC_{v_s}$): After receiving an SF message from parents, $v_s$ will check if it is an end node of a sibling edge candidate $(v_i, v_s)$. If it is, it will calculate the load $LdC_{v_s}$ diverted from $v_i$ to $v_j$ via the sibling edge $(v_i, v_s)$. Let $p_i$ be the number of parents of $v_s$. $LdC_{v_s}$ is equal to $Ld_{v_s}/p_s$.

- SF-s($v_s$, $v_i$, $v_j$, $LdRe_{v_s}$, $SL_{v_s}$): After receiving SF-c from sibling edge candidates, the base station chooses the node $v_s$ with the largest $LdC$ value. The energy level of the node is used to break the tie. Then, the base station sends an SF-s message to $v_s$. $v_s$ also uses SF-s to search for sibling edges. $LdRe_{v_s}$ is the remaining load that needs to be diverted from $v_i$ to $v_j$, and $SL_{v_s}$ is the number of sibling edges added to the current $k$-DAG.

- ADD-SIBLING($v_p$, $LdRe_{v_s}$, $SL_{v_s}$): This message is used to inform $v_p$ to search for a new sibling edge.

- SF-ACK($v_{sm}$, $v_i$, $v_j$, $LdRe_{v_{sm}}$, $SL_{v_{sm}}$): $v_{sm}$ sends SF-ACK message to the base station to indicate the end of the sibling edges search, and $LdRe_{v_{sm}}$ is the load not yet diverted from $v_i$ to $v_j$.

**Algorithm 3**: Adding sibling edges

**For the base station** $v_0$:

*exit=false*

*k-value=0 /\* if k-value is equal to k, no more sibling edges can be added to the k-DAG \*/*

**while** *exit=false* **do**

    find the child $v_i$ with the heaviest load such that at least one sibling edge is added for $v_i$ in the last round

    **if** such a $v_i$ does not exists **then**

        *exit=true*

        **exit while**

    **end if**

    find the child $v_j$ which is a neighbor of $v_i$ with the lightest load

    $LdBl_{v_i} = (Ld_{v_i} - Ld_{v_j})/2$

    broadcast SF($v_i$, $v_j$, $LdBl_{v_i}$) to $v_i$ and all its descendants

    set timer T1

    $LdDo_{max}=0$

    **repeat**

        **if** SF-c($v_q$, $v_i$, $v_j$, $LdC_{v_q}$) is received from $v_i$'s descendant $v_q$ **then**

            **if** $LdC_{v_q} > LdDo_{max}$ **then**

                $LdDo_{max} = LdC_{v_q}$

                $v_s = v_q$

            **end if**

        **end if**

    **until** T1 expires

    $LdRe_{v_s} = LdBl_{v_i}$

    send SF-s($v_s$, $v_i$, $v_j$, $LdRe_{v_s}$, $k$ - $k$-value) to $v_s$

    **loop**

        **if** SF-ACK($v_{sm}$, $v_i$, $v_j$, $LdRe_{v_{sm}}$, $SL_{v_{sm}}$) is received from $v_i$'s descendant $v_{sm}$ then

            $k$-value = $k$-value - $SL_{v_{sm}}$

            broadcast a message to each sensor node in the network to recalculate its load and the reachable base station

        **end if**

    **end loop**

**end while**

**For each sensor node** $v_s$:

**loop**

**if** SF($v_i$, $v_j$, $LdBl_{v_j}$) is received **then**

    **if** this SF message is not received before **then**

        broadcast SF($v_i$, $v_j$, $LdBl_{vi}$) to all its children

    **else**

        drop this SF message

    **end if**

    **if** $v_s$ is reachable from $v_j$ only via a sibling $v_t$ **then**

        calculate the load $LdC_{v_s}$ diverted from $v_i$ to $v_j$

        **if** $LdC_{v_s} < LdBl_{v_j}$

            send SF-c ($v_s$, $v_i$, $v_j$, $LdC_{v_s}$) to the base station

        **end if**

    **end if**

**else if** SF-s($v_s$, $v_i$, $v_j$, $LdRe_{vsi}$, $SL_{vsi}$) is received **then**

    add $v_i$ as the parent

    $LdRe_{v_s} = LdRe_{v_{si}} - LdC_{v_s}$

    $SL_{v_{si}} = SL_{v_{si}} + 1$

    **if** $v_s$ has a sibling $v_p$ and $SL_{v_s} < k$ **then**

        **if** $LdC_{v_p} \leq LdRe_{v_s}$ **then**

            send ADD-SIBLING($v_p$, $LdRe_{v_s}$, $SL_{v_s}$) to $v_p$

        **end if**

    **else if** $v_s$ is not a leaf node and $SL_{v_s} < k$ **then**

        send ADD-SIBLING($v_p$, $LdRe_{v_s}$, $SL_{v_s}$) to $v_p$

    **else**

        send SF-ACK($v_s$, $v_i$, $v_j$, $LdRe_{v_s}$, $SL_{v_s}$) to the base station

    **end if**

**else if** ADD-SIBLING($v_p$, $LdRe_{v_s}$, $SL_{v_s}$) is received **then**

    **if** ADD-SIBLING($v_p$, $LdRe_{v_s}$, $SL_{v_s}$) is received from a sibling **then**

        add $v_p$ as the parent

        $LdRe_{v_s} = LdRe_{v_{sl}} - LdC_{v_s}$

        $SL_{v_s} = SL_{v_{sl}} + 1$

    **end if**

    **if** $v_s$ has a sibling $v_p$ and $SL_{v_s} < k$ **then**

        **if** $LdC_{v_p} \leq LdRe_{v_s}$ **then**

            send ADD-SIBLING($v_s$, $LdRe_{v_s}$, $SL_{v_s}$) to $v_p$

        **end if**

    **else if** $v_s$ is not a leaf node and $SL_{v_s} < k$ **then**

       send ADD-SIBLING($v_p$, $LdRe_{v_s}$, $SL_{v_s}$) to $v_p$

    **else**

       send SF-ACK($v_s$, $v_t$, $v_j$, $LdRe_{v_s}$, $SL_{v_s}$) to the base station

   **end if**

  **end if**

**end loop**

Consider an example shown in Figure 3, where $S_x$ denotes a DAG rooted at a base station's child $v_x$. For simplicity, we assume that $k$ is equal to 2. First, the base station finds the child $v_i$ with the largest load, and the child $v_j$ with the smallest load among all its children that are also the neighbours of $v_i$. Next, the base station sends an SF message to all the sensor nodes in $S_i$. The only sibling edge is ($v_s$, $v_t$). Now, $v_s$ sends an SF-c message to the base station. After receiving the SF-c message from the only candidate $v_s$, the base station sends an SF-s message to $v_s$. Then, $v_s$ will add $v_t$ as its parent, i.e., adding the sibling edge ($v_s$, $v_t$) to the $k$-DAG. Next, $v_s$ sends an ADD-SIBLING($v_{si}$, $LdRe_{v_s}$, $SL_{v_s}$) message to its sibling $v_{si}$ to add the sibling edge ($v_s$,$v_{si}$) to the $k$-DAG. After that, $v_{si}$ sends ADD-SIBLING($v_{vsi}$, $LdRe_{vsi}$, $SL_{vsi}$) to its child $v_{sj}$. After receiving this message, $v_{sj}$ will not send this message to its child as no more sibling edge can be added to the $k$-DAG without violating the definition of the $k$-DAG. Therefore, $v_{sj}$ sends SF-ACK to the base station to indicate the completion of the current round of adding sibling edges. Lastly, the base station broadcast a message to each sensor node in the network to recalculate its load and the reachable base station children.

## 4. SIMULATION RESULTS AND ANALYSES

We evaluate our $k$-DAG based approach by comparing it with the SPD based approach proposed in [4]. We use lifetime and load balance as two metrics to evaluate the performance. We implement two routing algorithms, PE and MPE proposed in [4], on these two topologies, and compare the results of these two metrics.

A sensor node's lifetime depends on its energy consumption. As mentioned in [3], the major difference for energy consumption is from transmitting and receiving. So we ignore the energy consumption for listening, computing and sensing. The initial energy of each sensor node is 0.05 J energy. Each sensor node consumes 50 nanoJ for receiving 1 bit and 250 nanoJ for sending 1 bit [14], and all the sensor nodes generate data at the rate of 40 bits/hour. The hardware platform for our simulations is Intel Core i7 processor 2.3 GHz and 8 GB RAM.

As in [8], we use Chebyshev Sum Inequality as the criteria of load balance. Let $\{v_1, v_2... v_m\}$ be the set of the base station children, and $ld_{v_i}$ the load of a child $v_i$ of the base station. We use the following equation to calculate the balance factor $\theta$:

$$\theta = \frac{(\sum_{i=1}^{m} ld_{v_i})^2}{m\sum_{i=1}^{m} ld_{v_i}^2}$$

We use Cooja simulator to generate network instances, ignoring those instances with disconnected sensor nodes in the network. The transmission range for each sensor node is fixed to 50 *unit* in radius. All the sensor nodes are randomly deployed in a square area, from 100×100 to 350×350 *unit²* by increasing 50×50 *unit²* each time. The network size increases from 50 to 100 nodes by an increment of 10. There are a total of 6 scenarios with different network sizes, and 10 instances for each scenario, resulting in a total of 60 different network instances.

We calculate the balance factor by the data flow collected from the simulation results. The simulation results for average, maximum and minimum balance factors are shown in Figure 4, Figure 5 and Figure 6, respectively. In each figure, the horizontal axis indicates the number of sensor nodes, and the vertical axis represents the balance factor. For all the instances, *k*-DAG outperforms SPD by achieving higher load balance. The load balance improvements range from 0.1% to 83%. The largest increase of 83% occurs in a scenario with 50 sensor nodes.

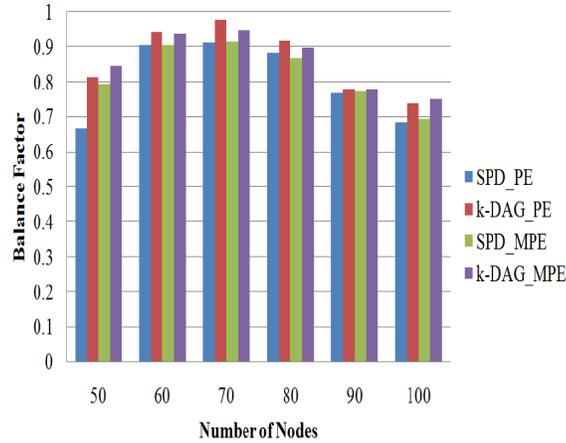

Figure 4. Average balance factors

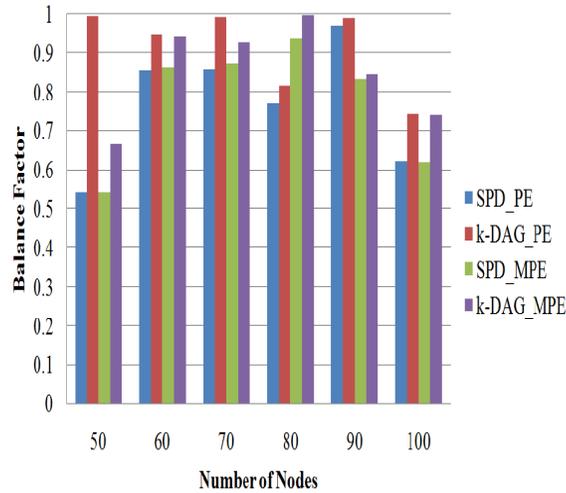

Figure 5. Maximum balance factors

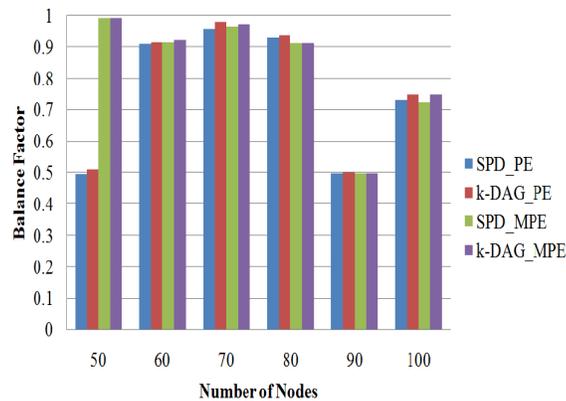

Figure 6. Minimum balance factors

Figure 7, Figure 8 and Figure 9 show the lifetimes for different scenarios. It can be seen that the lifetime of a WSN is not always inversely proportional to the number of sensor nodes of the WSN. Figure 8 shows the maximum lifetimes for each scenario. In a 50 nodes scenario, our approach achieves a maximum improvement of 82% for the network lifetime. It shows in Figure 9 that there is an instance for 50 nodes scenario with no improvement in network lifetime. However, *k*-DAG outperforms SPD for all the other instances.

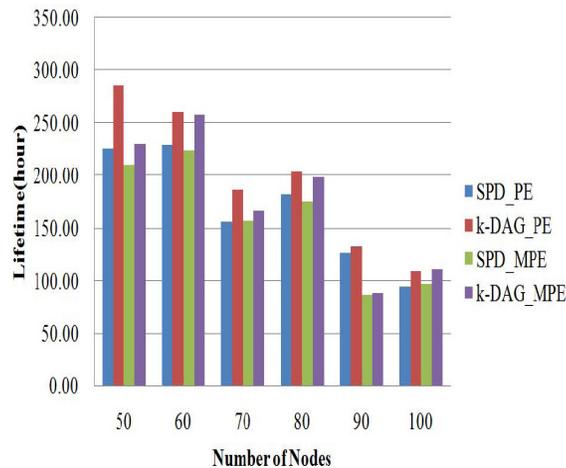

Figure 7. Average lifetimes

We choose one instance for each scenario to demonstrate the relationship between *k* and the network lifetime. In Figure 10, Max(p) is the longest path length in a *k-DAG*. The horizontal axis indicates the number of sensor nodes. The vertical axis denotes the ratio in percentage of the lifetime achieved by a particular k and the lifetime achieved by the maximum value of *k*. For 70 nodes scenario, adding the first 10% sibling edges, compared with Max(p), achieves 93.7% of the maximum lifetime. However, for a 50 nodes scenario, adding the first 10% sibling edges just achieves 15.3% of the maximum lifetime, and it improves to 70.6% when 30% sibling edges are added. It can be seen from the figure that the network lifetime is not linearly proportional to *k*. The network topology is a key factor affecting the impact of k on the network lifetime.

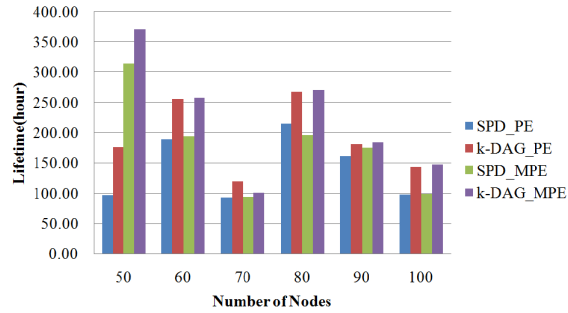

Figure 8. Maximum lifetimes

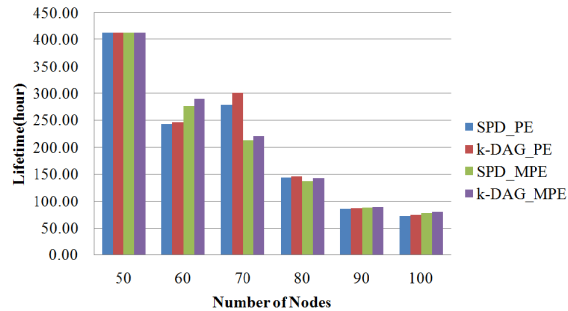

Figure 9. Minimum lifetimes

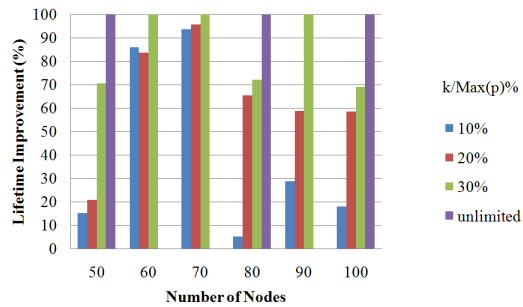

Figure 10. Lifetime versus *k*

From the simulation results, an instance with a longer network lifetime has a larger balance factor. However, a large balance factor does not guarantee a long lifetime. The number of children of the base station also has a significant impact on the network. For most instances of the same size, the more children the base station has, the longer network lifetime is achieved by our approach.

We also observe that in the instances with a small balance factor, the *k*-DAG significantly improves the network lifetime after only a few sibling edges are added into the *k*-DAG. The key reason is that the sibling edges connecting disjoint subgraphs greatly divert the load from the sensor nodes with heavy load to the sensor nodes with lighter load. Furthermore, the sibling edges at a higher level divert significantly more load than those at a lower level. In some instances, the algorithm does not optimize the network lifetime but just improves the balance factor. It occurs when the sensor node with the heaviest load among all the base station children cannot find a sibling edge to divert the load to other base station children with light load, but there are still sibling edges that can be added to the *k*-DAG for those base station children with light load. In these cases, the balance factor can be improved without lifetime increase.

## 5. CONCLUSION

In this paper, we study the problem of lifetime aware data collection in WSNs using DAG topology. We propose a *k*-DAG based approach which not only increases the lifetime of a WSN but also provides the maximum latency guarantee for data collection. We build a *k*-DAG in a distributed way. The *k*-DAG based approach achieves better load balance among the children of the base station to prolong the network lifetime. Meanwhile, it guarantees that the length of any path from each sensor node to the base station and the shortest path length differ by at most *k*. We have simulated our approach and compared it with the SPD based one by using a set of network instances and two routing algorithms. The simulation results show that our approach significantly outperforms the SPD based one in both network lifetime and load balance.

**Authors**

Jingjing Fei is currently a Master of Engineering by research candidate in School of Computer Science and Engineering, The University of New South Wales, Australia. He received a Master of Engineering Degree in Software Engineering from Huazhong University of Science and Technology (2007) and Bachelor of Science Degree in Computer Science and Technology from Wuhan University of Technology (2004). His current research interests focus on wireless sensor networks.

Hui Wu received PhD from National University of Singapore, ME and BE from Huazhong University of Science and Technology. His early career was mainly focused on CNC systems. He was the chief software architect and developer of Aerospace I CNC System, and a co-founder of Wuhan Huazhong Numerical Control Co. Ltd. His current research areas include embedded systems, parallel and distributed systems, and wireless sensor networks.

Yongixn Wang is currently a Master of Engineering by research candidate in Faculty of Engineering and Information Technology, University of Technology, Sydney, Australia. She received a Master of Engineering Degree in Software Engineering from Huazhong University of Science and Technology (2007) and Bachelor of Science Degree in Electrical and Electronic Engineering from Wuhan University of Science and Technology (2004). Her current research interests focus on packet scheduling in radio resource management for the future wireless networks